\documentclass{appolb}
\usepackage{graphicx}

\def\eq#1{{Eq.~(\ref{#1})}}
\def\fig#1{{Fig.~\ref{#1}}}
\newcommand{\as}{\alpha_s}

\begin{document}
\title{Brief Review of Saturation Physics%
  \thanks{Based on the lectures presented at the LIV$^{\mbox{th}}$
    Cracow School of Theoretical Physics {\sl QCD meets experiment} in
    Zakopane, Poland in June 2014 and on the overview talk given at
    the {\sl QCD Evolution Workshop} in Santa Fe, NM in May 2014.}%
}
\author{Yuri V. Kovchegov 
\address{Department of Physics, The Ohio State University,
  Columbus, OH 43210, USA}
}
\maketitle
\begin{abstract}
  We present a short overview of saturation physics followed by a
  summary of the recent progress in our understanding of nonlinear
  small-$x$ evolution. Topics include McLerran--Venugopalan model,
  Glauber--Mueller approximation, nonlinear BK/JIMWLK evolution
  equations, along with the running-coupling and NLO corrections to
  these equations. We conclude with selected topics in saturation
  phenomenology.
\end{abstract}


\section{Introduction}

\begin{figure}[htb]
\centerline{%
\includegraphics[width=0.6 \textwidth]{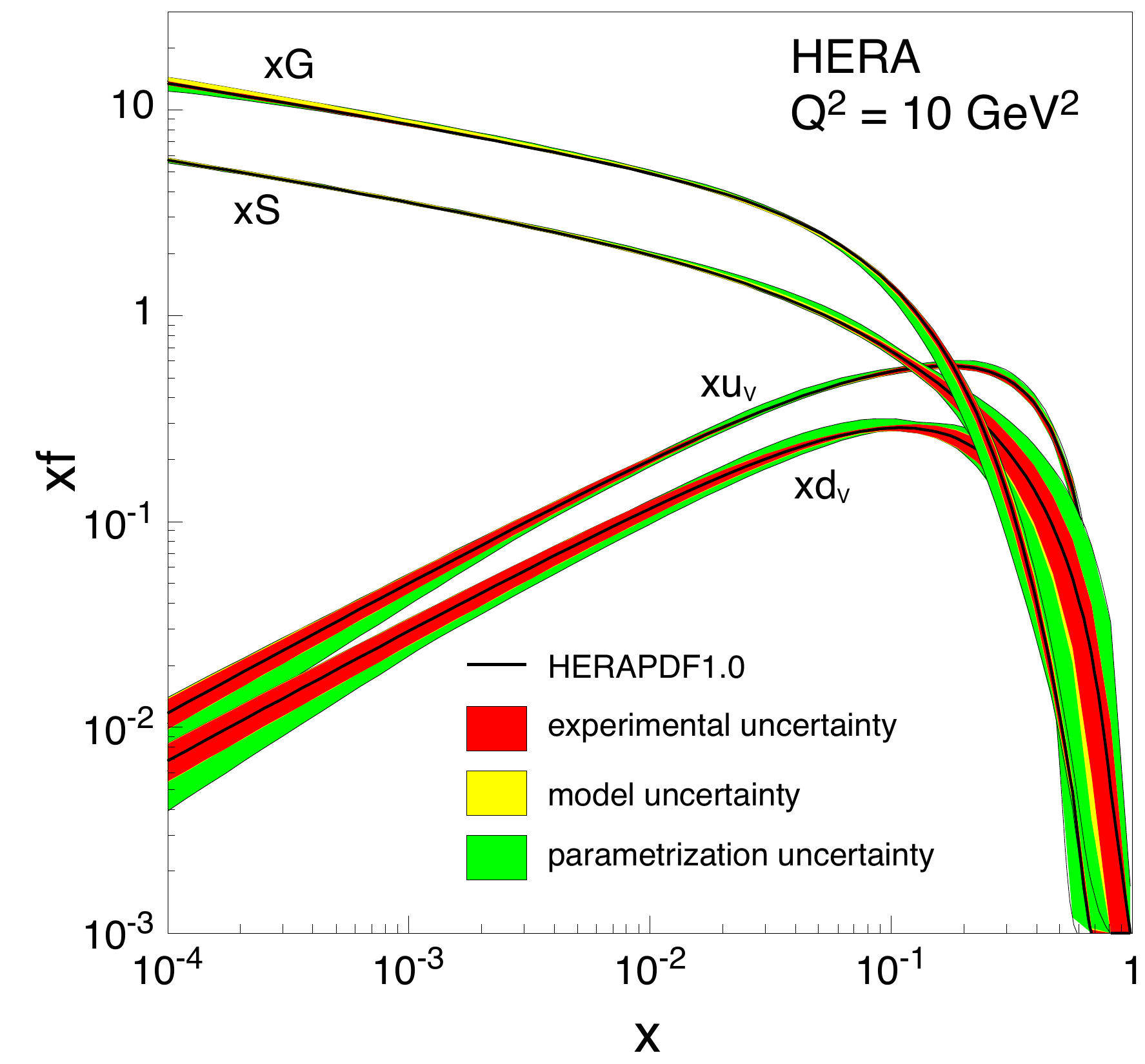}}
\caption{Parton distribution functions (PDFs) of a proton at the scale
  $Q^2 = 10$~GeV$^2$ plotted as functions of Bjorken $x$. Here $xu_v$
  and $xd_v$ are the valence quark distributions, $xS$ is the sea
  quark distribution, and $xG$ is the gluon distribution. Note that
  the vertical axis is logarithmic.}
\label{Fig:PDFs}
\end{figure}

Saturation physics is built on an observation that small Bjorken-$x$
part of the wave function for an ultrarelativistic hadron or nucleus
contains an intrinsic hard scale, {\sl the saturation scale} $Q_s$
\cite{Gribov:1984tu}. This scale characterizes the typical size of
color charge density fluctuations in the small-$x$ wave functions
\cite{McLerran:1993ka,McLerran:1993ni,McLerran:1994vd}. The number of
partons in the proton or nuclear wave function grows at small-$x$, as
shown in \fig{Fig:PDFs}, leading to a high density of quarks and
gluons inside the proton. This high density leads to large color
charge fluctuations, and, therefore, to a large value of $Q_s$.

In fact, a detailed calculation shows that the saturation scale grows
as
\begin{equation}
\label{Qs-scaling}
  Q_s^2 \sim A^{1/3} \, \left( \frac{1}{x} \right)^\lambda ,
\end{equation} 
where $A$ is the atomic number of the nucleus. The numerical value of
the inverse power of Bjorken-$x$ is approximately $\lambda \approx 0.2
- 0.3$. From \eq{Qs-scaling} we conclude that at small enough value of
$x$ and/or for large enough nucleus the saturation scale $Q_s$ becomes
larger than the QCD confinement scale $\Lambda_{QCD}$, $Q_s \gg
\Lambda_{QCD}$ such that the strong coupling constant becomes small,
\begin{equation}
  \label{eq:coupling}
  \as (Q_s^2) \ll 1. 
\end{equation}
Therefore, in the saturation regime we are dealing with a high density
of gluons and quarks inside the proton or nucleus, while at the same
time having a small coupling constant justifying the use of
perturbative expansion in the powers of $\as$.


\section{Classical Gluon Fields}

\begin{figure}[htb]
\centerline{%
\includegraphics[width=0.85 \textwidth]{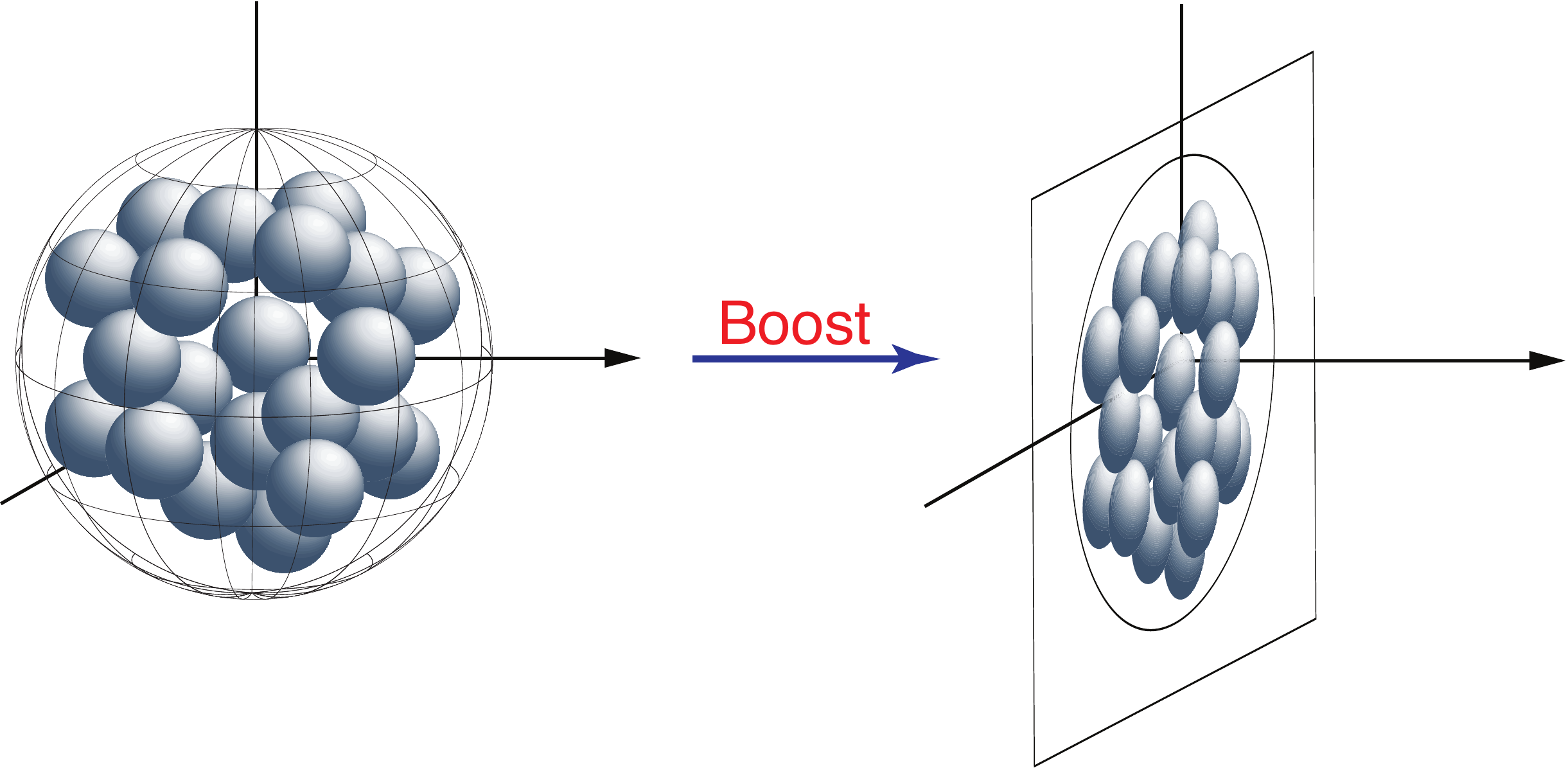}}
\caption{An ultrarelativistic nucleus appears as a ``pancake'' with
  the $A^{1/3}$-enhanced color charge density.}
\label{Fig:A-enhancement}
\end{figure}

The most convenient system to study saturation dynamics appears to be
the small-$x$ wave function of a large nucleus. From now one we will
concentrate on gluons, since they dominate over quarks at small-$x$ as
follows from \fig{Fig:PDFs}. The small-$x$ gluons ``see'' the whole
nucleus coherently in the longitudinal direction, and can be emitted
by any of the nucleons at a given impact parameter. (Note that a gluon
with $k_T \gg \Lambda_{QCD}$ is localized in the transverse coordinate
space and does not interact with the nucleons at other impact
parameters.) The small-$x$ gluon can originate in any of the $\sim
A^{1/3}$ nucleons at a given transverse position. If the nucleus is
ultrarelativistic this means that the gluon is emitted by the
effective color charge density which is enhanced by a factor of
$A^{1/3}$ compared to that in a single proton. This is illustrated in
\fig{Fig:A-enhancement}.

If we define the saturation scale squared as the gluon density in the
transverse plane, one readily obtains $Q_s^2 \sim A^{1/3}$, such that
for a large nucleus $Q_s \gg \Lambda_{QCD}$ and $\as (Q_s^2) \ll
1$. At small coupling the leading gluon field is classical (since one
can neglect quantum loop corrections): hence, to find the gluon field
of a nucleus one has to solve classical Yang-Mills equations 
\begin{equation}
  \label{eq:YM}
  {\cal D}_\mu F^{\mu\nu} = J^\nu
\end{equation}
with the nucleus providing the source current $J^\nu$. This is the
main concept behind the McLerran--Venugopalan model
\cite{McLerran:1993ka,McLerran:1993ni,McLerran:1994vd}.

\begin{figure}[htb]
\centerline{%
\includegraphics[width=0.75 \textwidth]{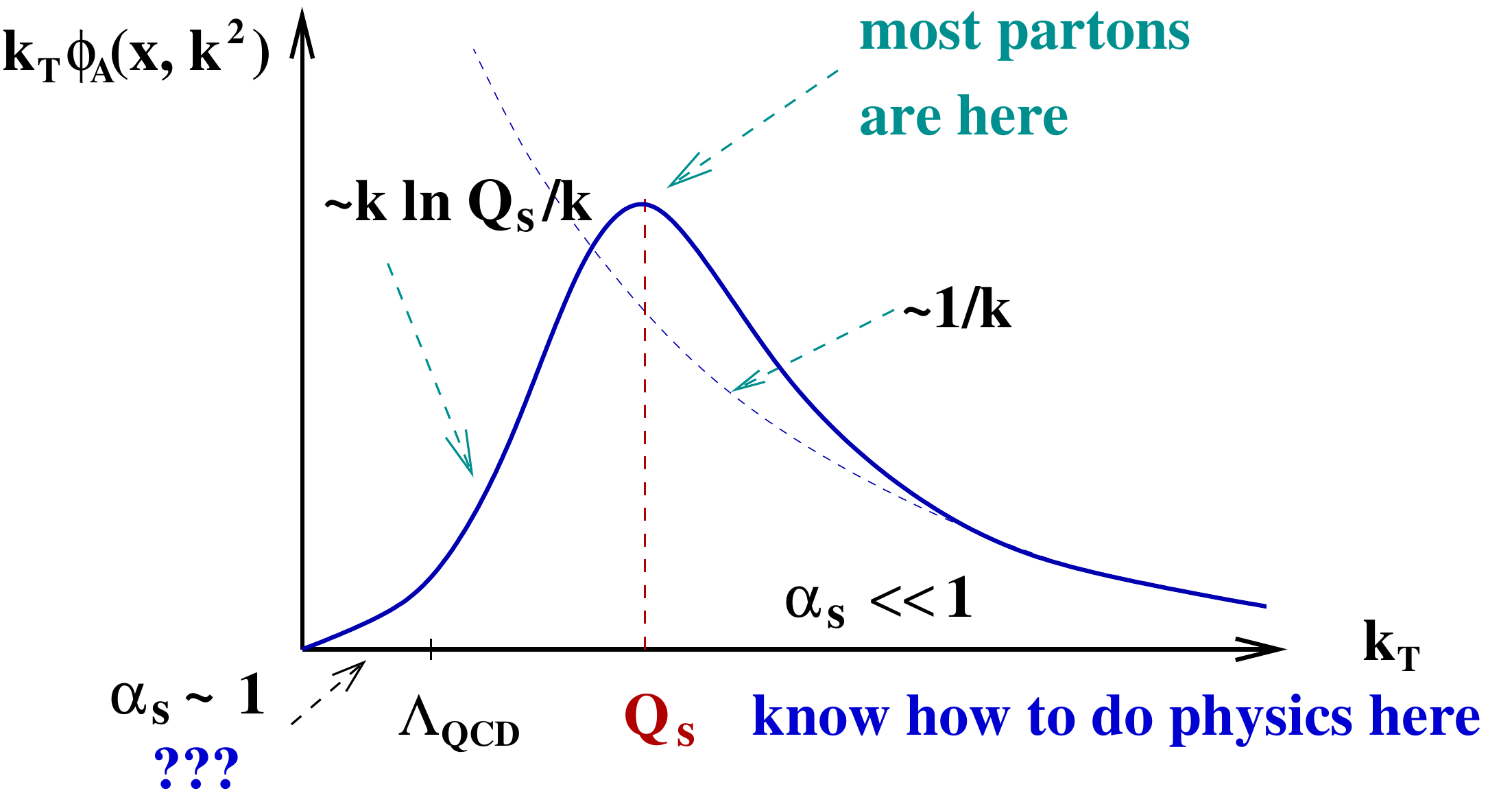}}
\caption{Unintegrated gluon distribution multiplied by the transverse
  momentum phase-space factor $k_T$ sketched as a function of $k_T$
  (solid line). The dashed line denotes the lowest-order result,
  without saturation effects.}
\label{Fig:unint_glue}
\end{figure}

The Yang-Mills equations (\ref{eq:YM}) were solved for a single
nucleus source in \cite{Kovchegov:1996ty,Jalilian-Marian:1997xn}.  The
resulting gluon field could be used to construct the unintegrated
gluon distribution of a nucleus $\phi_A (x, k_T^2)$, which counts the
number of gluons at a given values of Bjorken $x$ and transverse
momentum $k_T$:
\begin{equation}
  \label{eq:WW}
  \phi_A (x, k_T^2) =  \frac{C_F}{\as \, 2 \, \pi^3} \,
  \int d^2 b_\perp \, d^2 r_\perp \, e^{i {\bf k} \cdot
    {\bf r}} \frac{1}{r_\perp^2} \, \left[ 1 - 
    e^{- \frac{1}{4} \, r_\perp^2 \, Q_{s}^2 ({\vec b}_\perp) \, \ln (1/r_\perp
      \, \Lambda)} \right].
\end{equation}
Here the gluon saturation scale is given by 
\begin{equation}
  \label{eq:QsG}
  Q_{s}^2 ({\bf b}) \, = \, 4 \, \pi \, \as^2 \, T ({\bf b})
\end{equation}
with $T ({\bf b})$ the nuclear profile function. Transverse vectors
are denoted by ${\bf x} = (x^1, x^2)$ and $x_\perp = x_T = |{\bf
  x}|$. The unintegrated gluon distribution (gluon TMD) $\phi_A (x,
k_T^2)$, multiplied by the transverse momentum phase-space factor of
$k_T$ is plotted schematically in \fig{Fig:unint_glue} as a function
of $k_T$. We conclude from this plot that the majority of gluons in
this classical nuclear wave function have transverse momentum $k_T
\sim Q_s \gg \Lambda_{QCD}$, such that applicability of perturbation
theory is justified.

\begin{figure}[htb]
\centerline{%
\includegraphics[width=0.85 \textwidth]{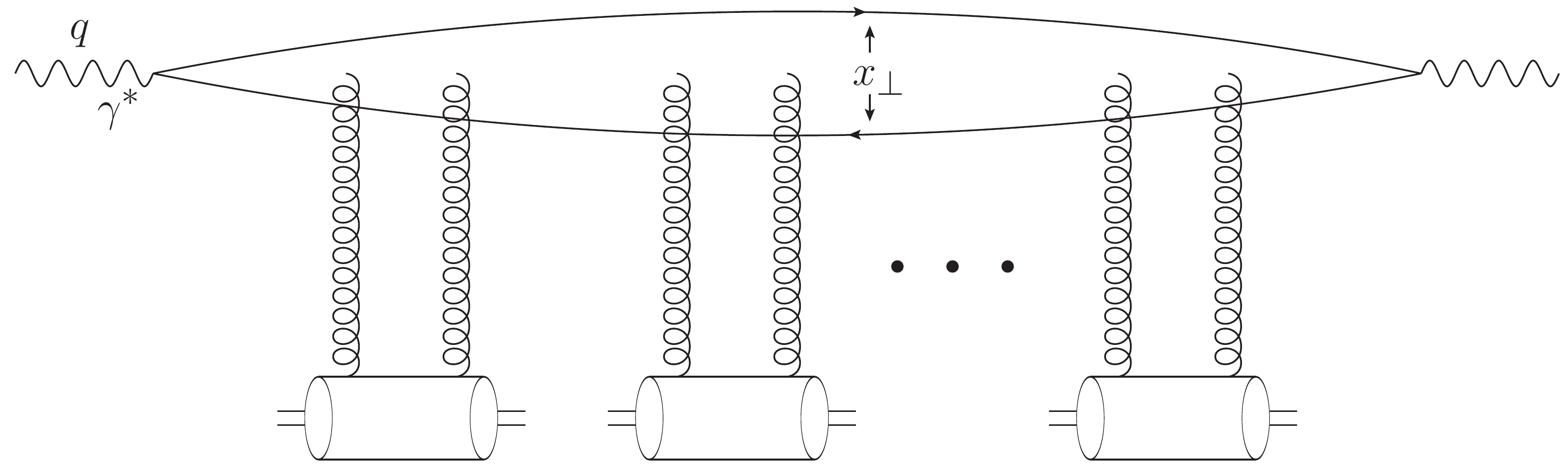}}
\caption{DIS on a large nucleus in the Glauber--Mueller
  approximation.}
\label{Fig:DIS}
\end{figure}

Now let us consider deep inelastic scattering (DIS) on a large
nucleus, working in the same classical approximation. The DIS process
at high energies is shown in \fig{Fig:DIS}: the electron (not shown)
emits a virtual photon, which then splits into a $q\bar q$ pair which
scatters on the nuclear target. At the lowest-order each interaction
with the nucleons in the nucleus is limited to a two-gluon exchange:
this is known as the Glauber--Mueller model \cite{Mueller:1989st}. The
resummation parameter of such approximation is then $\as^2 \,
A^{1/3}$. It can be shown that this is also the parameter resummed by
the classical gluon fields in the MV model \cite{Kovchegov:1997pc}.

The high-energy DIS cross section can be written as a convolution of
the light-cone wave function $\Psi^{\gamma^* \rightarrow q {\bar q}}$
of the virtual photon splitting into a $q\bar q$ pair and the
scattering amplitude of the $q\bar q$ on the nuclear target,
\begin{equation}
  \label{eq:dipole}
  \sigma_{tot}^{\gamma^* A} (x, Q^2) \, = \, \int \frac{d^2 x_\perp}{4
    \, \pi} \, \int\limits_0^1 \, \frac{dz}{z \, (1-z)} \ 
  |\Psi^{\gamma^* \rightarrow q {\bar q}} ({\bf x}, z)|^2 \ 
  \sigma_{tot}^{q {\bar q} A} ({\bf x}, Y),
\end{equation}
where $Y = \ln 1/x$ is the rapidity variable, $x_\perp$ is the
transverse size of the dipole, and $z$ is the fraction of the virtual
photon's light-cone momentum carried by the quark.

\begin{figure}[htb]
\centerline{%
\includegraphics[width=0.5 \textwidth]{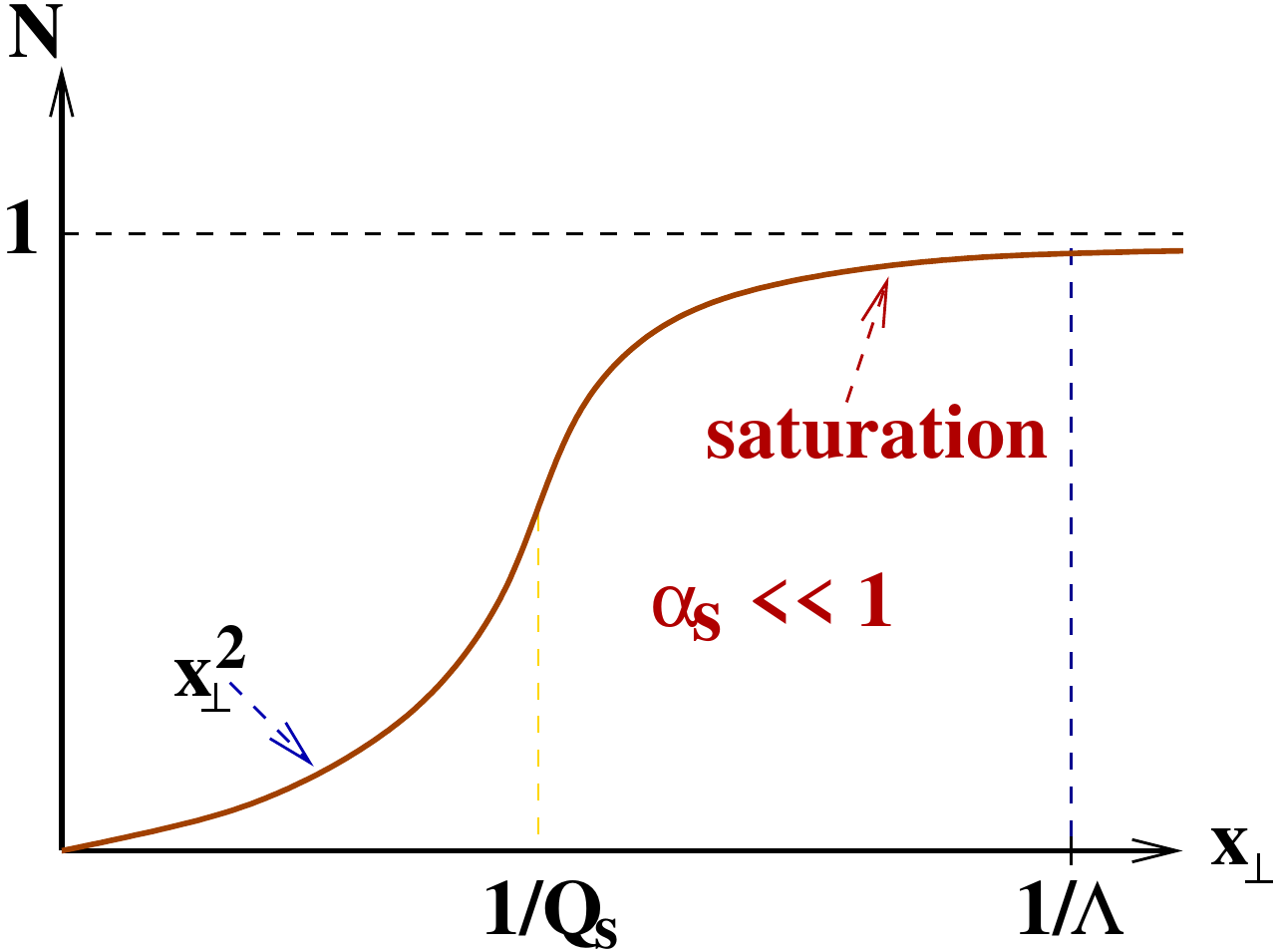}}
\caption{Dipole--nucleus forward scattering amplitude as a function of
  the dipole size $x_\perp$ in the Glauber--Mueller model.}
\label{Fig:GM}
\end{figure}

One can write the dipole--nucleus cross section as an integral over
impact parameters of the (imaginary part of the) forward
dipole--nucleus scattering amplitude $N$,
\begin{equation}
  \label{eq:dipAA}
  \sigma_{tot}^{q {\bar q} A} ({\bf x}, Y) \, = \, 2 \, \int
  d^2 b \ N ({\bf x}, {\bf b}, Y).
\end{equation}
The dipole--nucleus forward scattering amplitude is found in the
Glauber--Mueller model to be \cite{Mueller:1989st}
\begin{equation}
  \label{glaN2}
  N({\bf x}, {\bf b} , Y=0) \, = \, 1 - \exp \left\{ -
    \frac{x_\perp^2 \, Q_{sq}^2 ({\bf b}) \, \ln (1/x_\perp \,
      \Lambda)}{4} \, \right\}
\end{equation}
with the quark saturation scale
\begin{equation}
  \label{qsmv}
  Q_{sq}^2 ({\bf b}) \, \equiv \, \frac{4 \, \pi \, \as^2 \,
    C_F}{N_c} \, T({\bf b}). 
\end{equation}

The amplitude $N$ from \eq{glaN2} is sketched as a function of the
dipole size $x_\perp$ in \fig{Fig:GM}. As the dipole size goes to
zero, so does the amplitude $N$. This is the manifestation of color
transparency: zero-size dipole does not interact. As the dipole size
increases, so does the amplitude $N$ again. However, due to multiple
rescatterings effects of \fig{Fig:DIS} which led to the exponentiation
in \eq{glaN2}, we always have $N < 1$. Using this bound in
\eq{eq:dipAA} we see that, for a nucleus of radius $R$ it translates
into $\sigma_{tot}^{q {\bar q} A} < 2 \pi R^2$, which is the
well-known black disk limit. We see that saturation effects lead to
the scattering cross section that preserves the black disk limit: we
can invert this observation to argue that saturation is a consequence
of unitarity. Note that the onset of saturation effects and the
approach to the black disk regime happens around $x_\perp \sim 1/Q_s$,
where the dipole is still perturbatively small and perturbation theory
is applicable.


\section{Small-$x$ Evolution}

\begin{figure}[htb]
\centerline{%
\includegraphics[width=0.8 \textwidth]{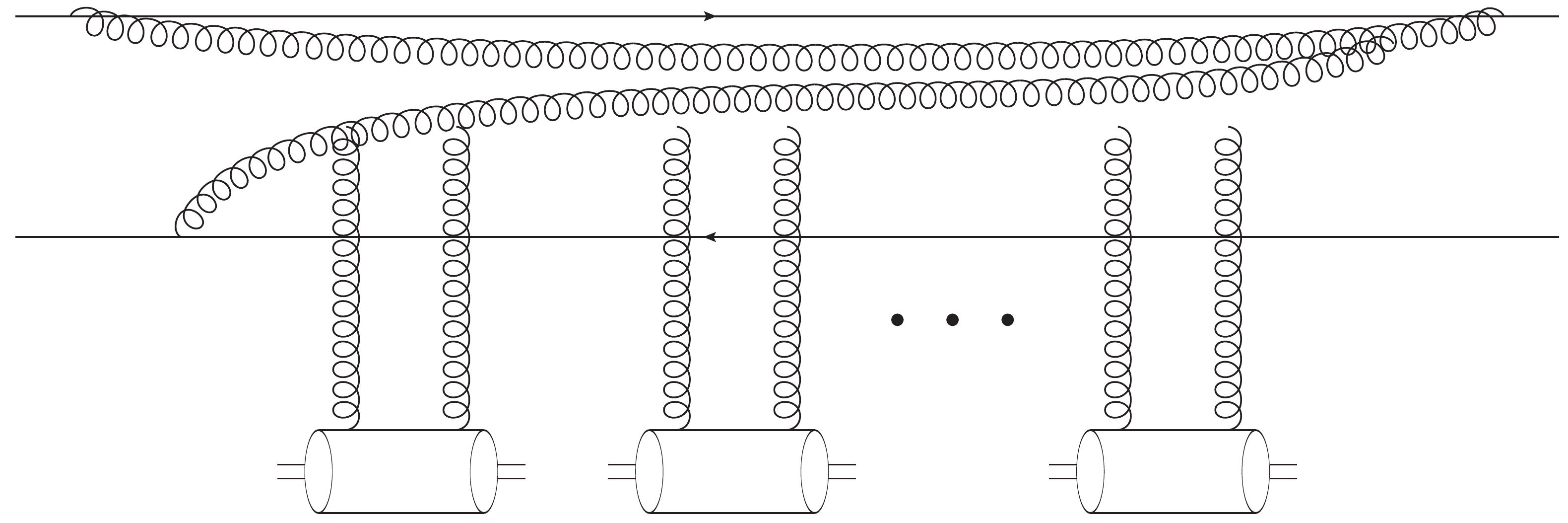}}
\caption{Small-$x$ evolution corrections to the dipole--nucleus
  forward scattering amplitude.}
\label{Fig:evolution}
\end{figure}

The classical picture presented above lacks the energy (or
Bjorken-$x$) dependence. The energy-dependence enters the picture
through quantum evolution corrections. The corrections for
dipole--nucleus scattering in DIS are illustrated in
\fig{Fig:evolution}. Each gluon emission brings in a power of $\as$
and, due to the phase-space integral, a factor of rapidity $Y$ (at the
leading order). The resulting leading-logarithmic approximation (LLA)
resums powers of $\as \, Y$.

Small-$x$ evolution corrections in the large-$N_c$ approximation can
be absorbed into the dipole amplitude $N$ with the help of the
Balitsky--Kovchegov (BK) evolution equation
\cite{Balitsky:1996ub,Balitsky:1998ya,Kovchegov:1999yj,Kovchegov:1999ua},
\begin{eqnarray}
  \frac{\partial}{\partial Y} \, N_{{\bf x}_{1}, {\bf x}_{0}} (Y ) \, 
& = & \, \frac{\as \, N_c}{2 \, \pi^2} \, \int \, d^2
  x_2 \, \frac{x^2_{10}}{x^2_{20}\,x^2_{21}} \, \bigg[ N_{{\bf
      x}_{1}, {\bf x}_{2}} (Y) + N_{{\bf
      x}_{2}, {\bf x}_{0}} (Y) \nonumber \\ & - & N_{{\bf x}_{1},
    {\bf x}_{0}}  (Y) - N_{{\bf x}_{1}, {\bf x}_{2}} (Y) \, N_{{\bf x}_{2}, {\bf x}_{0}} 
(Y) \bigg].
\label{BK}
\end{eqnarray}
We have slightly modified our notation: the dipole--nucleus amplitude,
now denoted by $N_{{\bf x}_{1}, {\bf x}_{0}} (Y )$, depends on the
positions of the quark and the anti-quark (${\bf x}_{1}, {\bf x}_{0}$)
in the dipole. Above $x_{ij} = |{\bf x}_i - {\bf x}_j|$. The initial
condition for \eq{BK} is given by the Glauber--Mueller formula
(\ref{glaN2}): this way one resums both the powers of $\as \, Y$ and
$\as^2 \, A^{1/3}$. The linear terms on the right-hand side of \eq{BK}
correspond to the Balitsky--Fadin--Kuraev--Lipatov (BFKL) evolution
equation \cite{Kuraev:1977fs,Balitsky:1978ic}, while the quadratic
term introduced damping due to saturation effects.

No closed integro-differential equation for the amplitude $N$ exists
beyond the large-$N_c$ approximation. Instead, for general-$N_c$ one
has to solve the
Jalilian-Marian--Iancu--McLerran--Weigert--Leonidov--Kovner (JIMWLK) \\
evolution equation
\cite{Jalilian-Marian:1997dw,Jalilian-Marian:1997gr,Iancu:2001ad,Iancu:2000hn},
which is a functional differential equation, giving energy dependence
not only for the dipole operator, but for any other operator made out
of eikonal Wilson lines along the light-cone. Interestingly enough, a
numerical solution of both the BK and JIMWLK evolution equations in
\cite{Rummukainen:2003ns,Kovchegov:2008mk} indicates that the
differences between the large-$N_c$ and any-$N_c$ expressions for the
dipole amplitude $N$ are very small, on the order of $0.1 \%$, much
smaller than the naively anticipated $1/N_c^2 \approx 0.1$.

\begin{figure}[htb]
\centerline{%
\includegraphics[width=0.5 \textwidth]{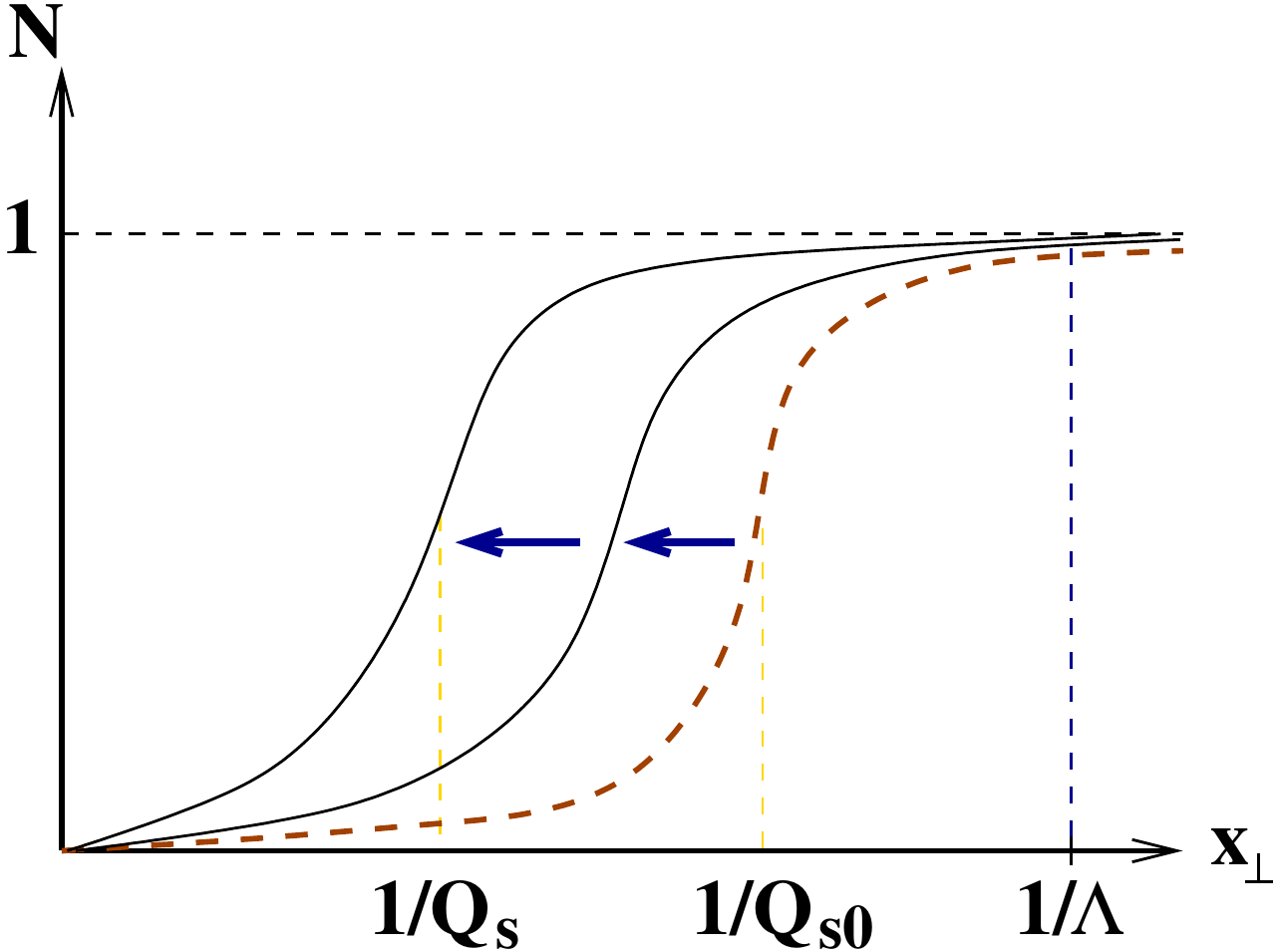}}
\caption{A sketch of the small-$x$ evolution of the dipole amplitude
  $N$: the initial condition (given by the Glauber-Mueller formula) is
  shown by the dashed line. As rapidity increases the dipole amplitude
  shifts to the smaller values of $x_\perp$ in the plot, as indicated
  by the solid-line curves and the arrows.}
\label{Fig:BKsol}
\end{figure}

No exact analytic solution of \eq{BK} exists. Our understanding of its
solution stems from several approximate analytic solutions
\cite{Gribov:1984tu,Iancu:2002tr,Mueller:2002zm,Levin:1999mw} along
with the exact numerical solutions
\cite{Golec-Biernat:2001if,Braun:2000wr,Levin:2001et,Albacete:2004gw}. Qualitative
behavior of the solution of \eq{BK} is shown in \fig{Fig:BKsol}. There
we see that small-$x$ evolution makes the dipole amplitude shift left
from the initial conditions (dashed curve), toward the smaller values
of the dipole size $x_\perp$. The two evolved curves are shown by
solid lines, with the direction of rapidity increase denoted by
arrows. We see two important features in this solution.  One is that
we always have $N<1$: indeed $N=1$ is the fixed point of the evolution
(\ref{BK}), such that the black disk limit is always preserved by the
nonlinear evolution. Hence nonlinear small-$x$ evolution is {\sl
  unitary}! Another feature is that the saturation scale, as the
characteristic of the transition into the saturation region, is
growing with rapidity: in \fig{Fig:BKsol} we clearly have $Q_s >
Q_{s0}$ with $Q_{s0}$ the initial value of the saturation scale. A
more careful analysis leads to $Q_s^2 \sim (1/x)^\lambda \sim
e^{\lambda \, Y}$ scaling of the saturation scale with decreasing
Bjorken $x$ or increasing rapidity, justifying the claim we made in
\eq{Qs-scaling}. The solution of the BK/JIMWLK evolution for the
dipole amplitude has another important property known as the geometric
scaling \cite{Stasto:2000er}: the dipole amplitude turns out to be a
function of only one variable, $N (x_\perp, Y) = N (x_\perp \, Q_s
(Y))$, over a broad range of the dipole sizes $x_\perp$
\cite{Gribov:1984tu,Iancu:2002tr,Levin:1999mw}.

\begin{figure}[htb]
\centerline{%
\includegraphics[width=0.5 \textwidth]{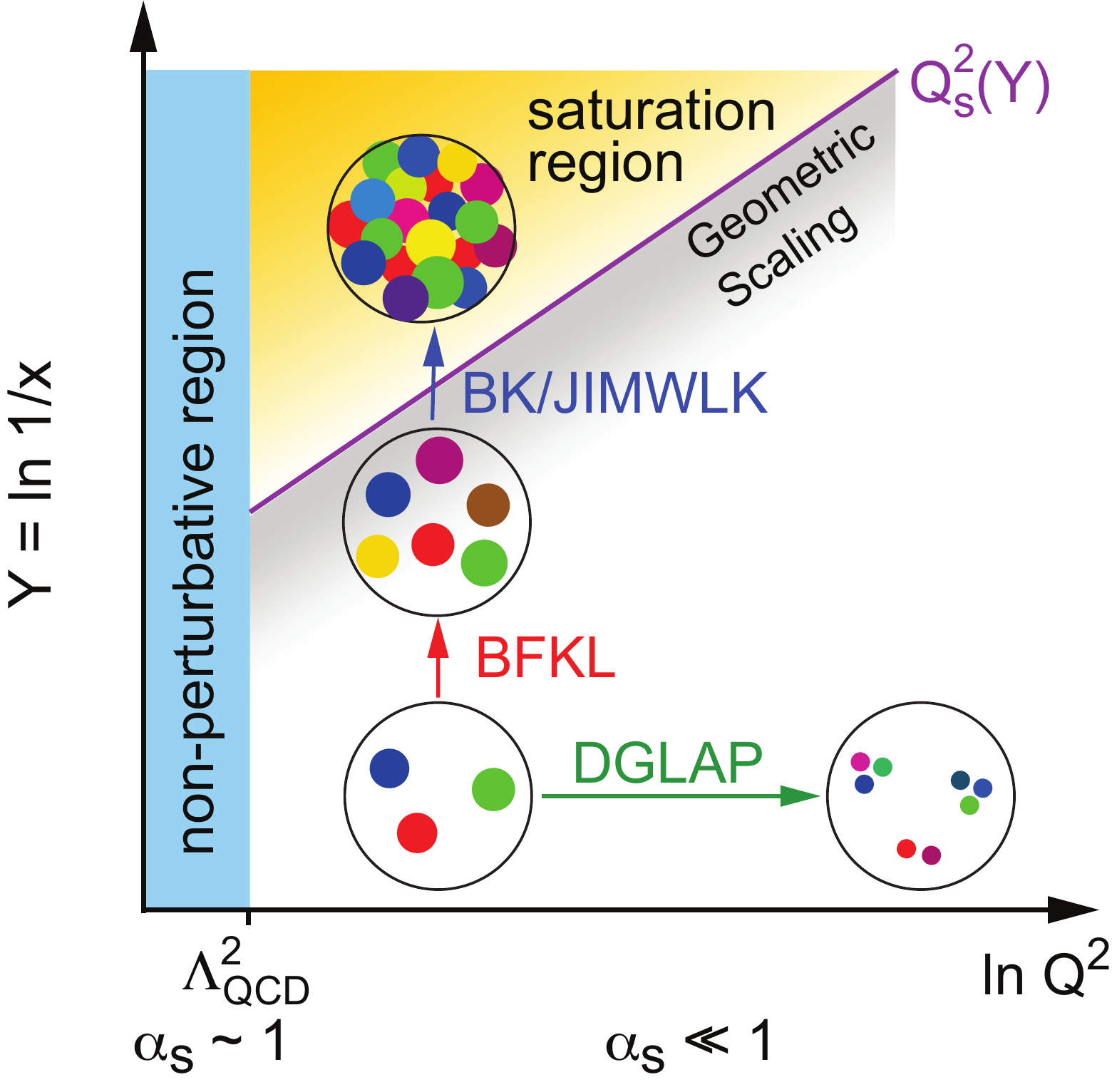}}
\caption{A map of high-energy QCD.}
\label{Fig:Map}
\end{figure}

We summarize this discussion of the nonlinear small-$x$ evolution with
a map of high-energy QCD in \fig{Fig:Map}. 
There we plot the action of the QCD evolution equations in the $(\ln
Q^2, \ln 1/x)$ plane. The DGLAP evolution equation, implementing
renormalization group flow, evolves PDF toward large $Q^2$ with
approximately fixed values of $x$. The linear BFKL evolution equation
evolves the unintegrated gluon distribution (or dipole amplitude)
toward low-$x$, but eventually stops being applicable due to violation
of unitarity. The nonlinear BK/JIMWLK equations take over the BFKL
evolution at low-$x$ preventing the unitarity violation and guiding
the system into the saturation region.

For much more detailed presentations of saturation physics we
recommend the review articles
\cite{Gribov:1984tu,Iancu:2003xm,Jalilian-Marian:2005jf,Weigert:2005us,Gelis:2010nm,Albacete:2014fwa}
along with the book \cite{KovchegovLevin}.


\section{Higher-Order Corrections to the BK and JIMWLK Evolution
  Equations}

Over the past decade, main progress in our understanding of nonlinear
small-$x$ evolution came from calculation of higher-order corrections
to it, and from successful phenomenological applications of the
results of those calculations.

The first development in this direction was the calculation of the
running-coupling correction to the BK and JIMWLK evolution equations
in \cite{Gardi:2006rp,Balitsky:2006wa,Kovchegov:2006vj} using the BLM
method \cite{BLM}. The result, in the scheme used in
\cite{Kovchegov:2006vj}, reads
\begin{eqnarray}\label{eqNrc}
  && \frac{\partial N_{{\bf x}_{0}, {\bf x}_1} (Y)}{\partial Y} \, = \,
  \frac{N_c}{2 \pi^2} \, \int d^2 x_2  \nonumber \\  && \times \left[ \as
    \left(\frac{1}{x_{20}^2}\right) \,
    \frac{1}{x_{20}^2} - 2 \, \frac{\as \left(\frac{1}{x_{20}^2}\right) 
      \ \as \left( \frac{1}{x_{21}^2} \right)}{ \as \left( 
        \frac{1}{R^2} \right)} \, \frac{{\bf x}_{20} \cdot {\bf
        x}_{21}}{x_{20}^2 \, x_{21}^2} 
    + \as \left(\frac{1}{x_{21}^2}\right) \, \frac{1}{x_{21}^2}
  \right] \nonumber \\ &&
  \times \, \left[ N_{{\bf x}_{0}, {\bf x}_2} (Y) + N_{{\bf x}_{2},
      {\bf x}_1} (Y) - N_{{\bf x}_{0}, {\bf x}_1} (Y) - N_{{\bf x}_{0},
      {\bf x}_2} (Y) \, N_{{\bf x}_{2}, {\bf x}_1} (Y) \right],
\end{eqnarray}
with the scale $R$ given by
\begin{eqnarray}\label{Rexp}
  R^2 ({\bf x}_0, {\bf x}_1 ; {\bf z}) &&  = \, |{\bf z} -{\bf x}_0| \,
  |{\bf z}-{\bf x}_1| \nonumber \\ \times && \! \! \! \! 
 \left( \frac{|{\bf z}-{\bf x}_1|}{|{\bf z}
      -{\bf x}_0|} \right)^{\frac{({\bf z} -{\bf x}_0)^2 + ({\bf
        z}-{\bf x}_1)^2}{({\bf z} - {\bf x}_0)^2 - ({\bf z}-{\bf
        x}_1)^2} - 2 \, \frac{|{\bf z}-{\bf x}_0|^2 \, |{\bf z}-{\bf
        x}_1|^2}{({\bf z}-{\bf x}_0) \cdot ({\bf z}-{\bf x}_1)} \,
    \frac{1}{|{\bf z} -{\bf x}_0|^2 - |{\bf z}-{\bf x}_1|^2}}.
\end{eqnarray}
We will refer to the BK evolution equation with running coupling
corrections as rcBK. The effect of the running-coupling corrections
on the small-$x$ evolution is to suppress the contribution from the
very small dipoles (due to asymptotic freedom), thus slowing down the
evolution. In fact the parameter $\lambda$ in \eq{Qs-scaling} goes
from being about $0.7 - 0.8$ at fixed QCD coupling down to about $0.2
- 0.3$ when the running coupling corrections are included
\cite{Albacete:2004gw,Albacete:2007yr,Albacete:2007sm}: this is a
positive development, since $\lambda \approx 0.2 - 0.3$ gives us the
energy dependence close to that observed in experimental data.

More recently, the full next-to-leading order (NLO) correction to the
BK evolution kernel was calculated in a formidable calculation
presented in \cite{Balitsky:2008zz}. (Running-coupling evolution of
\eq{eqNrc} contains a subset of NLO and higher-order corrections, but
is not a complete NLO or higher-order result.) Knowledge of NLO BK
corrections is an important part of theoretical progress in the
field. Solution of the NLO BK evolution (analytical or, more likely,
numerical) has not been constructed at the time of writing.

NLO corrections for the JIMWLK kernel were obtained very recently in
\cite{Grabovsky:2013gta,Balitsky:2013fea,Kovner:2013ona}. Similar to
BK evolution, the impact of the NLO JIMWLK corrections on the
evolution of Wilson line correlators is yet to be determined.

NLO correction to the BK or JIMWLK evolution kernel is
order-$\as^2$. If one solves NLO BK/JIMWLK evolution equation exactly,
one would be resumming powers of $\as^2 \, Y$, in addition to the
powers of $\as \, Y$ resummed to all orders by the LLA evolution. Here
one runs into the standard power-counting conundrum: two iterations of
NLO evolution kernel give a contribution of the order $(\as^2 \,
Y)^2$, which is of the same order as one iteration of the
leading-order (LO) kernel times an iteration of the
next-to-next-to-leading order (NNLO) kernel, $(\as \, Y) \, (\as^3 \,
Y)$. It is thus {\it a priori} not clear whether construction of an
all-order solution of the NLO non-linear evolution equation is
parametrically justified, or whether this would overstep the precision
of the approximation. Perhaps knowledge of the overall structure of
the solution would facilitate this perturbative expansion (e.g. in
DGLAP evolution all perturbative expansion resides in one place in the
solution --- the anomalous dimensions): while such program has
recently been initiated for the linear BFKL evolution
\cite{Chirilli:2013kca}, it would be much harder to do for the
nonlinear evolution case, where we do not know the exact analytic
solution even in the LLA.


\begin{figure}[hbt]
\centerline{%
\includegraphics[width=0.7 \textwidth]{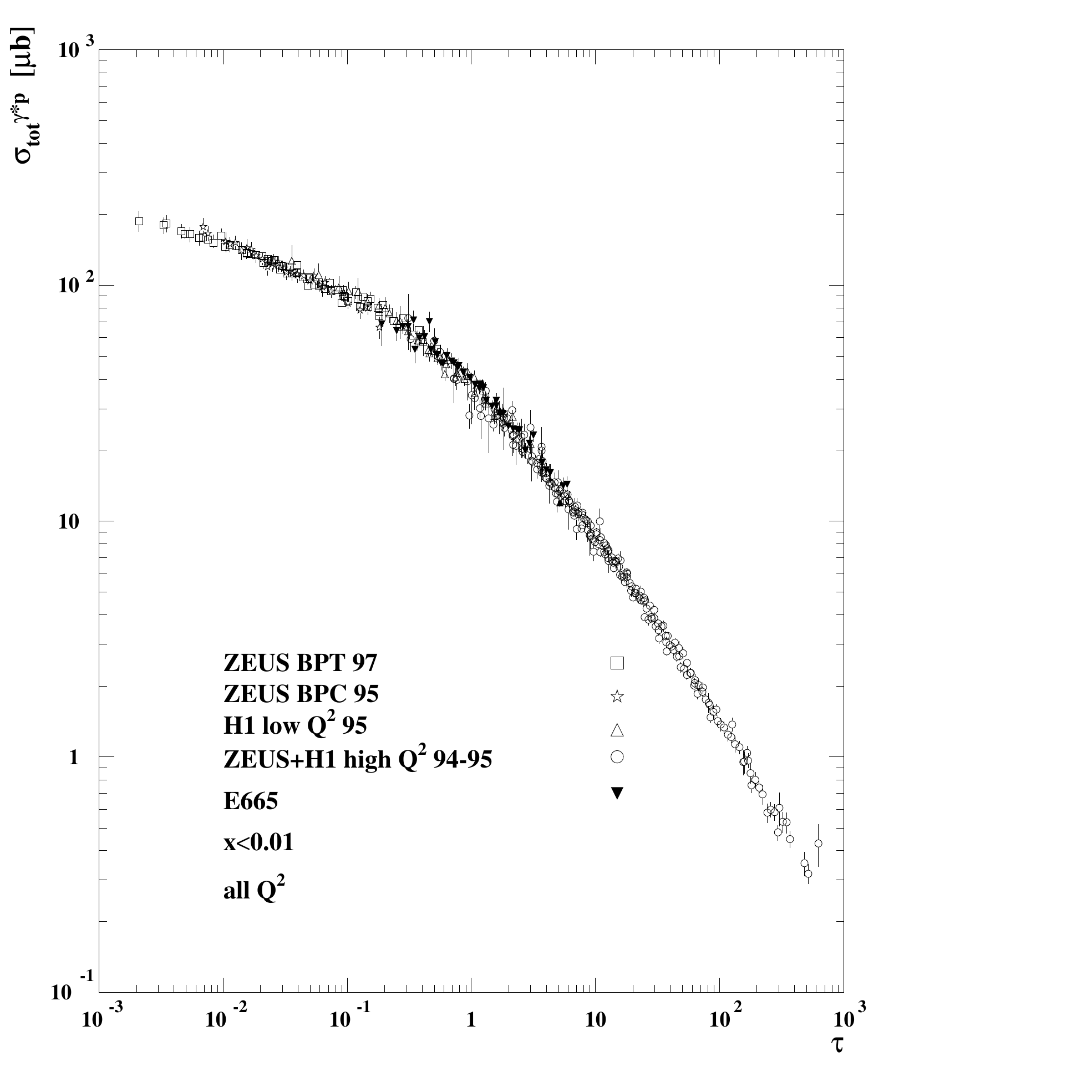}}
\caption{Data on DIS $\gamma^* + p$ total cross section for $x<0.01$
  plotted as a function of $\tau = Q^2/Q_s^2 (x)$
  \cite{Stasto:2000er}.}
\label{Fig:scaling}
\end{figure}

\section{Some Saturation Phenomenology}

The field of phenomenological applications of saturation physics has
grown tremendously over the last decade, encompassing scattering
processes as diverse as DIS, $p+p$, $p+A$ and heavy ion collisions
(see \cite{Albacete:2014fwa} for an up-to-date review of saturation
phenomenology). It is impossible to do full justice to this area in
this short review. Instead we will only present a few phenomenological
successes of saturation physics.

As discussed above, geometric scaling is a consequence of non-linear
small-$x$ evolution. In \fig{Fig:scaling} from \cite{Stasto:2000er} we
show a compilation of DIS total cross section data for $x<10^{-2}$
plotted as a function of the single scaling variable $\tau = Q^2/Q_s^2
(x)$. The figure demonstrates that small-$x$ DIS data appears to
exhibit geometric scaling predicted by saturation theory
\cite{Gribov:1984tu,Iancu:2002tr,Levin:1999mw}!

\begin{figure}[htb]
\centerline{%
\includegraphics[width=0.7 \textwidth]{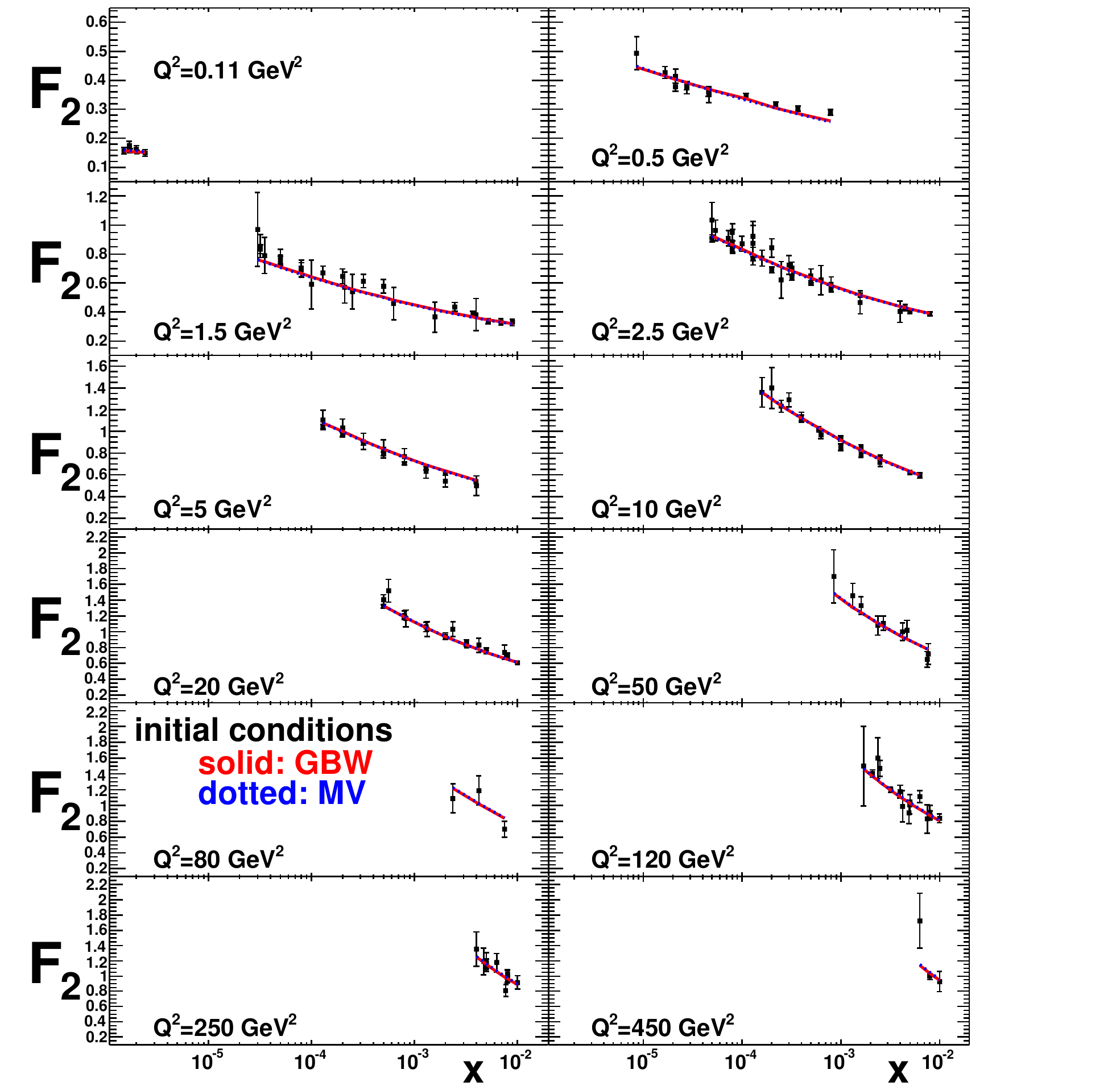}}
\caption{Fit to the HERA DIS data on $F_2$ structure function of the
  proton at low-$x$ within the saturation framework performed in
  \cite{Albacete:2009fh} based on the rcBK evolution equation.}
\label{Fig:F2}
\end{figure}

A more quantitative comparison of small-$x$ DIS data and saturation
theory from \cite{Albacete:2009fh,Albacete:2010sy} is shown in
\fig{Fig:F2} for the $F_2$ structure function of the proton. The
theory curves shown are generated using rcBK evolution equation
(employing equations like (\ref{eq:dipole}) and (\ref{eq:dipAA}) to
obtain the structure function). Clearly the description of the data is
very good.

Saturation physics and nonlinear small-$x$ evolution are relevant not
only to DIS, but to any high-energy scattering process. They are
likely to play an important role in describing particle production and
correlations originating in the early stages of heavy ion
collisions. Description of heavy ion collisions in the saturation
framework starts with determining classical gluon field of the
colliding ions in the MV model. One has to solve the same \eq{eq:YM},
but now with the source given by {\sl two} colliding nuclei. This
problem is very hard to solve analytically, allowing only for either
perturbative or variational solutions
\cite{Kovner:1995ja,Kovchegov:1997ke,Kovchegov:1998bi,Kovchegov:2000hz,Balitsky:2004rr,Blaizot:2010kh}. Luckily
the problem can be solved numerically
\cite{Krasnitz:1999wc,Krasnitz:2003nv,Lappi:2003bi}. Once the
classical gluon production is understood, one needs to include quantum
evolution corrections into the obtained formula: at present this is
impossible to do analytically, though it is doable numerically
\cite{Gelis:2008sz}. The program is similar to what was done for DIS:
quasi-classical Glauber--Mueller formula received quantum evolution
corrections through the BK/JIMWLK equations.

\begin{figure}[htb]
\centerline{%
\includegraphics[width=0.7 \textwidth]{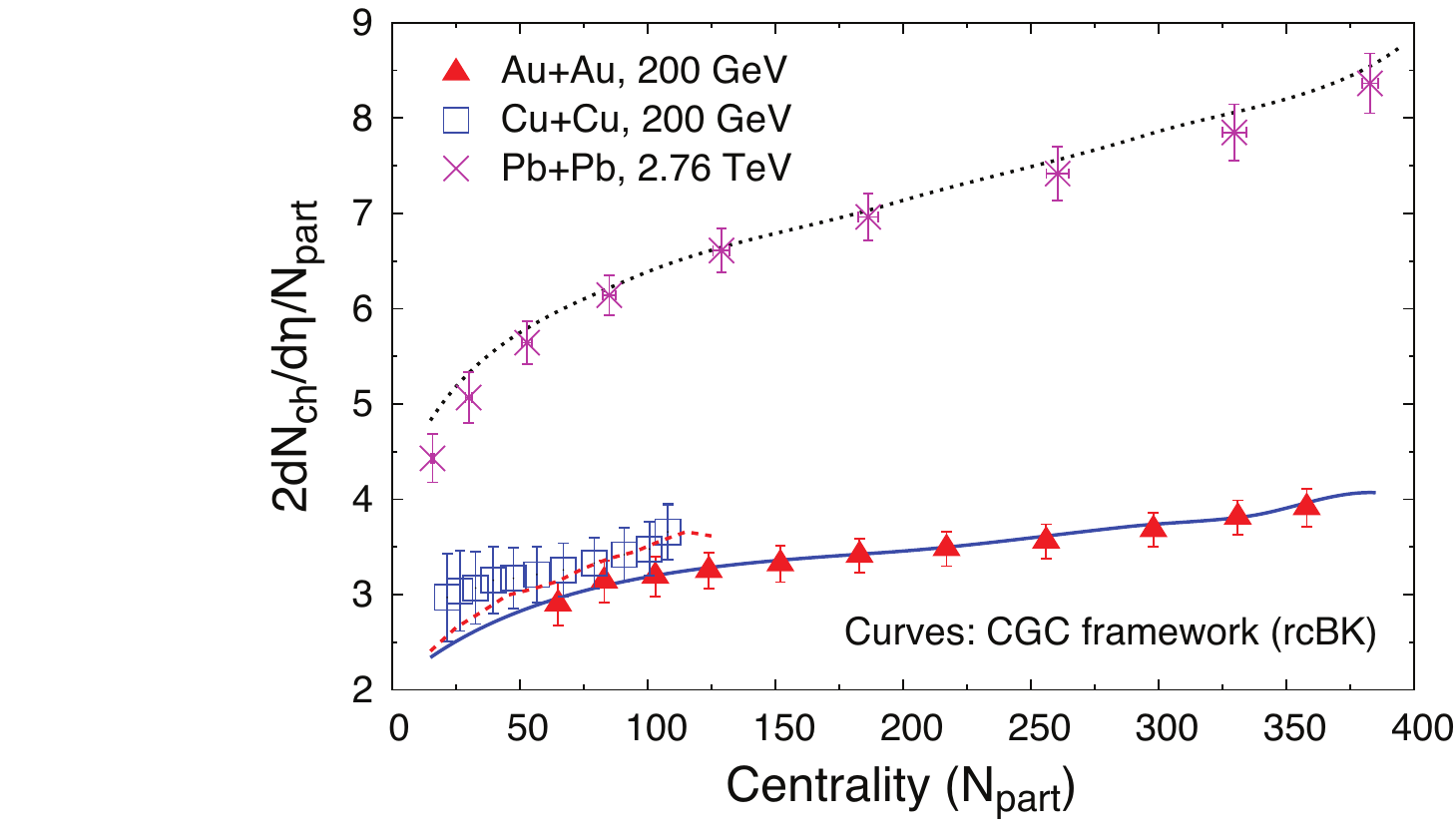}}
\caption{Saturation description of the charged hadron multiplicity in
  heavy ion collisions based on the KLN-type model with rcBK evolution
  \cite{ALbacete:2010ad}. RHIC data (Au+Au and Cu+Cu) was fitted
  (solid and lower dashed lines), while the LHC results (Pb+Pb) were
  {\sl predicted} by the top dotted curve.}
\label{Fig:AAmult}
\end{figure}

While the exact saturation calculation for gluon production in heavy
ion collisions is rather hard to do, one could use an approximate
Kharzeev--Levin--Nardi (KLN) approach
\cite{Kharzeev:2001gp,Kharzeev:2000ph,Kharzeev:2007zt} which employs
the (slightly modified) $k_T$-factorization formula which is an exact
saturation-physics result for gluon production in $p+A$ collisions
\cite{Kovchegov:2001sc,Kharzeev:2003wz}. The unintegrated gluon
distributions entering the $k_T$-factorization formula can be found
using rcBK-evolved expression for the dipole amplitude. The result of
applying this procedure to charged hadron production in $A+A$
collisions is shown in \fig{Fig:AAmult}: while the description of RHIC
data in this figure is a result of a fit, the dashed curve for the LHC
is a prediction, appearing to be in a very good agreement with the
data. Once again saturation physics appears to be consistent with the
data, and this time it was in fact able to predict the data.


\section{Outlook}

While tantalizing evidence for saturation regime was seen in $e+p$,
$p+A$ and $A+A$ collisions, the decisive evidence for saturation
sealing the discovery case can be found at an $e+A$ collider. In
high-energy $e+A$ collisions the saturation scale (\ref{Qs-scaling})
would get enhancements from both the low value of $x$ and the large
value of $A$, making the saturation region much broader than in $e+p$
collisions. Another advantage of $e+A$ collisions is a clean electron
probe, allowing for higher precision in theoretical predictions and,
with varying virtuality of the photon $Q^2$, providing an extra lever
for experimental measurements, giving $e+A$ collisions an advantage
over $p+A$ and $A+A$ collisions in terms of its potential for
saturation discovery. An Electron Ion Collider (EIC) is being proposed
in the US: for more details on the proposal we refer the reader to the
EIC White Paper \cite{Accardi:2012qut}.


\section*{Acknowledgments}

I would like to thank the organizers of the LIV$^{\mbox{th}}$ Cracow
School of Theoretical Physics in Zakopane, and in particular Michal
Praszalowicz, for hosting such an enjoyable meeting.

This material is based upon work supported by the
U.S. Department of Energy, Office of Science, Office of Nuclear
Physics under Award Number DE-SC0004286. \\


\begin{thebibliography}{10}

\bibitem{Gribov:1984tu}
L.~V. Gribov, E.~M. Levin, and M.~G. Ryskin, {\it {Semihard Processes in QCD}},
   {\em Phys. Rept.} {\bf 100} (1983) 1--150.

\bibitem{McLerran:1993ka}
L.~D. McLerran and R.~Venugopalan, {\it Gluon distribution functions for very
  large nuclei at small transverse momentum},  {\em Phys. Rev.} {\bf D49}
  (1994) 3352--3355, [\href{http://xxx.lanl.gov/abs/hep-ph/9311205}{{\tt
  hep-ph/9311205}}].

\bibitem{McLerran:1993ni}
L.~D. McLerran and R.~Venugopalan, {\it Computing quark and gluon distribution
  functions for very large nuclei},  {\em Phys. Rev.} {\bf D49} (1994)
  2233--2241, [\href{http://xxx.lanl.gov/abs/hep-ph/9309289}{{\tt
  hep-ph/9309289}}].

\bibitem{McLerran:1994vd}
L.~D. McLerran and R.~Venugopalan, {\it Green's functions in the color field of
  a large nucleus},  {\em Phys. Rev.} {\bf D50} (1994) 2225--2233,
  [\href{http://xxx.lanl.gov/abs/hep-ph/9402335}{{\tt hep-ph/9402335}}].

\bibitem{Kovchegov:1996ty}
Y.~V. Kovchegov, {\it Non-abelian {Weizs\"{a}cker-Williams} field and a two-
  dimensional effective color charge density for a very large nucleus},  {\em
  Phys. Rev.} {\bf D54} (1996) 5463--5469,
  [\href{http://xxx.lanl.gov/abs/hep-ph/9605446}{{\tt hep-ph/9605446}}].

\bibitem{Jalilian-Marian:1997xn}
J.~Jalilian-Marian, A.~Kovner, L.~D. McLerran, and H.~Weigert, {\it The
  intrinsic glue distribution at very small x},  {\em Phys. Rev.} {\bf D55}
  (1997) 5414--5428, [\href{http://xxx.lanl.gov/abs/hep-ph/9606337}{{\tt
  hep-ph/9606337}}].

\bibitem{Mueller:1989st}
A.~H. Mueller, {\it {Small x Behavior and Parton Saturation: A QCD Model}},
  {\em Nucl. Phys.} {\bf B335} (1990) 115.

\bibitem{Kovchegov:1997pc}
Y.~V. Kovchegov, {\it {Quantum structure of the non-Abelian
  Weizs\"{a}cker-Williams field for a very large nucleus}},  {\em Phys. Rev.}
  {\bf D55} (1997) 5445--5455,
  [\href{http://xxx.lanl.gov/abs/hep-ph/9701229}{{\tt hep-ph/9701229}}].

\bibitem{Balitsky:1996ub}
I.~Balitsky, {\it Operator expansion for high-energy scattering},  {\em Nucl.
  Phys.} {\bf B463} (1996) 99--160,
  [\href{http://xxx.lanl.gov/abs/hep-ph/9509348}{{\tt hep-ph/9509348}}].

\bibitem{Balitsky:1998ya}
I.~Balitsky, {\it Factorization and high-energy effective action},  {\em Phys.
  Rev.} {\bf D60} (1999) 014020,
  [\href{http://xxx.lanl.gov/abs/hep-ph/9812311}{{\tt hep-ph/9812311}}].

\bibitem{Kovchegov:1999yj}
Y.~V. Kovchegov, {\it Small-x {$F_2$} structure function of a nucleus including
  multiple pomeron exchanges},  {\em Phys. Rev.} {\bf D60} (1999) 034008,
  [\href{http://xxx.lanl.gov/abs/hep-ph/9901281}{{\tt hep-ph/9901281}}].

\bibitem{Kovchegov:1999ua}
Y.~V. Kovchegov, {\it Unitarization of the {BFKL} pomeron on a nucleus},  {\em
  Phys. Rev.} {\bf D61} (2000) 074018,
  [\href{http://xxx.lanl.gov/abs/hep-ph/9905214}{{\tt hep-ph/9905214}}].

\bibitem{Kuraev:1977fs}
E.~A. Kuraev, L.~N. Lipatov, and V.~S. Fadin, {\it {The Pomeranchuk
  singularity in non-Abelian gauge theories}},  {\em Sov. Phys. JETP} {\bf 45}
  (1977) 199--204.

\bibitem{Balitsky:1978ic}
I.~Balitsky and L.~Lipatov, {\it {The Pomeranchuk Singularity in Quantum
  Chromodynamics}},  {\em Sov.J.Nucl.Phys.} {\bf 28} (1978) 822--829.

\bibitem{Jalilian-Marian:1997dw}
J.~Jalilian-Marian, A.~Kovner, and H.~Weigert, {\it The {Wilson}
  renormalization group for low x physics: Gluon evolution at finite parton
  density},  {\em Phys. Rev.} {\bf D59} (1998) 014015,
  [\href{http://xxx.lanl.gov/abs/hep-ph/9709432}{{\tt hep-ph/9709432}}].

\bibitem{Jalilian-Marian:1997gr}
J.~Jalilian-Marian, A.~Kovner, A.~Leonidov, and H.~Weigert, {\it The {Wilson}
  renormalization group for low x physics: Towards the high density regime},
  {\em Phys. Rev.} {\bf D59} (1998) 014014,
  [\href{http://xxx.lanl.gov/abs/hep-ph/9706377}{{\tt hep-ph/9706377}}].

\bibitem{Iancu:2001ad}
E.~Iancu, A.~Leonidov, and L.~D. McLerran, {\it {The renormalization group
  equation for the color glass condensate}},  {\em Phys. Lett.} {\bf B510}
  (2001) 133--144.

\bibitem{Iancu:2000hn}
E.~Iancu, A.~Leonidov, and L.~D. McLerran, {\it Nonlinear gluon evolution in
  the color glass condensate. {I}},  {\em Nucl. Phys.} {\bf A692} (2001)
  583--645, [\href{http://xxx.lanl.gov/abs/hep-ph/0011241}{{\tt
  hep-ph/0011241}}].

\bibitem{Rummukainen:2003ns}
K.~Rummukainen and H.~Weigert, {\it Universal features of {JIMWLK} and {BK}
  evolution at small $x$},  {\em Nucl. Phys.} {\bf A739} (2004) 183--226,
  [\href{http://xxx.lanl.gov/abs/hep-ph/0309306}{{\tt hep-ph/0309306}}].

\bibitem{Kovchegov:2008mk}
Y.~V. Kovchegov, J.~Kuokkanen, K.~Rummukainen, and H.~Weigert, {\it
  {Subleading-$N_c$ corrections in non-linear small-$x$ evolution}},  {\em
  Nucl. Phys.} {\bf A823} (2009) 47--82,
  [\href{http://xxx.lanl.gov/abs/0812.3238}{{\tt arXiv:0812.3238}}].

\bibitem{Iancu:2002tr}
E.~Iancu, K.~Itakura, and L.~McLerran, {\it Geometric scaling above the
  saturation scale},  {\em Nucl. Phys.} {\bf A708} (2002) 327--352,
  [\href{http://xxx.lanl.gov/abs/hep-ph/0203137}{{\tt hep-ph/0203137}}].

\bibitem{Mueller:2002zm}
A.~H. Mueller and D.~N. Triantafyllopoulos, {\it The energy dependence of the
  saturation momentum},  {\em Nucl. Phys.} {\bf B640} (2002) 331--350,
  [\href{http://xxx.lanl.gov/abs/hep-ph/0205167}{{\tt hep-ph/0205167}}].

\bibitem{Levin:1999mw}
E.~Levin and K.~Tuchin, {\it Solution to the evolution equation for high parton
  density QCD},  {\em Nucl. Phys.} {\bf B573} (2000) 833--852,
  [\href{http://xxx.lanl.gov/abs/hep-ph/9908317}{{\tt hep-ph/9908317}}].

\bibitem{Golec-Biernat:2001if}
K.~Golec-Biernat, L.~Motyka, and A.~M. Stasto, {\it {Diffusion into infra-red
  and unitarization of the BFKL pomeron}},  {\em Phys. Rev.} {\bf D65} (2002)
  074037, [\href{http://xxx.lanl.gov/abs/hep-ph/0110325}{{\tt
  hep-ph/0110325}}].

\bibitem{Braun:2000wr}
M.~Braun, {\it Structure function of the nucleus in the perturbative {QCD} with
  ${N}_c \to \infty$ ({BFKL} pomeron fan diagrams)},  {\em Eur. Phys. J.} {\bf
  C16} (2000) 337--347, [\href{http://xxx.lanl.gov/abs/hep-ph/0001268}{{\tt
  hep-ph/0001268}}].

\bibitem{Levin:2001et}
E.~Levin and M.~Lublinsky, {\it Parton densities and saturation scale from
  non-linear evolution in dis on nuclei},  {\em Nucl. Phys.} {\bf A696} (2001)
  833--850, [\href{http://xxx.lanl.gov/abs/hep-ph/0104108}{{\tt
  hep-ph/0104108}}].

\bibitem{Albacete:2004gw}
J.~L. Albacete, N.~Armesto, J.~G. Milhano, C.~A. Salgado, and U.~A. Wiedemann,
  {\it {Numerical analysis of the Balitsky-Kovchegov equation with running
  coupling: Dependence of the saturation scale on nuclear size and rapidity}},
  {\em Phys. Rev.} {\bf D71} (2005) 014003,
  [\href{http://xxx.lanl.gov/abs/hep-ph/0408216}{{\tt hep-ph/0408216}}].

\bibitem{Stasto:2000er}
A.~M. Stasto, K.~Golec-Biernat, and J.~Kwiecinski, {\it Geometric scaling for
  the total $\gamma^* p$ cross-section in the low x region},  {\em Phys. Rev.
  Lett.} {\bf 86} (2001) 596--599,
  [\href{http://xxx.lanl.gov/abs/hep-ph/0007192}{{\tt hep-ph/0007192}}].

\bibitem{Iancu:2003xm}
E.~Iancu and R.~Venugopalan, {\it The color glass condensate and high energy
  scattering in {QCD}},  \href{http://xxx.lanl.gov/abs/hep-ph/0303204}{{\tt
  hep-ph/0303204}}.

\bibitem{Jalilian-Marian:2005jf}
J.~Jalilian-Marian and Y.~V. Kovchegov, {\it Saturation physics and deuteron
  gold collisions at {RHIC}},  {\em Prog. Part. Nucl. Phys.} {\bf 56} (2006)
  104--231, [\href{http://xxx.lanl.gov/abs/hep-ph/0505052}{{\tt
  hep-ph/0505052}}].

\bibitem{Weigert:2005us}
H.~Weigert, {\it Evolution at small {$x_{bj}$: The Color Glass Condensate}},
  {\em Prog. Part. Nucl. Phys.} {\bf 55} (2005) 461--565,
  [\href{http://xxx.lanl.gov/abs/hep-ph/0501087}{{\tt hep-ph/0501087}}].

\bibitem{Gelis:2010nm}
F.~Gelis, E.~Iancu, J.~Jalilian-Marian, and R.~Venugopalan, {\it {The Color
  Glass Condensate}},  {\em Ann.Rev.Nucl.Part.Sci.} {\bf 60} (2010) 463--489,
  [\href{http://xxx.lanl.gov/abs/1002.0333}{{\tt arXiv:1002.0333}}].

\bibitem{Albacete:2014fwa}
J.~L. Albacete and C.~Marquet, {\it {Gluon saturation and initial conditions
  for relativistic heavy ion collisions}},  {\em Prog.Part.Nucl.Phys.} {\bf 76}
  (2014) 1--42, [\href{http://xxx.lanl.gov/abs/1401.4866}{{\tt
  arXiv:1401.4866}}].

\bibitem{KovchegovLevin}
Y.~V. Kovchegov and E.~Levin, {\em Quantum Chromodynamics at High Energy}.
\newblock Cambridge University Press, 2012.

\bibitem{Gardi:2006rp}
E.~Gardi, J.~Kuokkanen, K.~Rummukainen, and H.~Weigert, {\it Running coupling
  and power corrections in nonlinear evolution at the high-energy limit},  {\em
  Nucl. Phys.} {\bf A784} (2007) 282--340,
  [\href{http://xxx.lanl.gov/abs/hep-ph/0609087}{{\tt hep-ph/0609087}}].

\bibitem{Balitsky:2006wa}
I.~I. Balitsky, {\it {Quark Contribution to the Small-$x$ Evolution of Color
  Dipole}},  {\em Phys. Rev. D} {\bf 75} (2007) 014001,
  [\href{http://xxx.lanl.gov/abs/hep-ph/0609105}{{\tt hep-ph/0609105}}].

\bibitem{Kovchegov:2006vj}
Y.~Kovchegov and H.~Weigert, {\it {Triumvirate of Running Couplings in
  Small-$x$ Evolution}},  {\em Nucl. Phys. {\bf A}} {\bf 784} (2007) 188--226,
  [\href{http://xxx.lanl.gov/abs/hep-ph/0609090}{{\tt hep-ph/0609090}}].

\bibitem{BLM}
S.~J. Brodsky, G.~P. Lepage, and P.~B. Mackenzie, {\it On the elimination of
  scale ambiguities in perturbative quantum chromodynamics},  {\em Phys. Rev.}
  {\bf D28} (1983) 228.

\bibitem{Albacete:2007yr}
J.~L. Albacete and Y.~V. Kovchegov, {\it Solving high energy evolution equation
  including running coupling corrections},  {\em Phys. Rev.} {\bf D75} (2007)
  125021, [\href{http://xxx.lanl.gov/abs/0704.0612}{{\tt 0704.0612}}].

\bibitem{Albacete:2007sm}
J.~L. Albacete, {\it {Particle multiplicities in Lead-Lead collisions at the
  LHC from non-linear evolution with running coupling}},  {\em Phys. Rev.
  Lett.} {\bf 99} (2007) 262301, [\href{http://xxx.lanl.gov/abs/0707.2545}{{\tt
  0707.2545}}].

\bibitem{Balitsky:2008zz}
I.~Balitsky and G.~A. Chirilli, {\it {Next-to-leading order evolution of color
  dipoles}},  {\em Phys. Rev.} {\bf D77} (2008) 014019,
  [\href{http://xxx.lanl.gov/abs/0710.4330}{{\tt arXiv:0710.4330}}].

\bibitem{Grabovsky:2013gta}
A.~Grabovsky, {\it {On the solution to the NLO forward BFKL equation}},  {\em
  JHEP} {\bf 1309} (2013) 098, [\href{http://xxx.lanl.gov/abs/1307.3152}{{\tt
  arXiv:1307.3152}}].

\bibitem{Balitsky:2013fea}
I.~Balitsky and G.~A. Chirilli, {\it {Rapidity evolution of Wilson lines at the
  next-to-leading order}},  {\em Phys.Rev.} {\bf D88} (2013) 111501,
  [\href{http://xxx.lanl.gov/abs/1309.7644}{{\tt arXiv:1309.7644}}].

\bibitem{Kovner:2013ona}
A.~Kovner, M.~Lublinsky, and Y.~Mulian, {\it {Complete JIMWLK Evolution at
  NLO}},  \href{http://xxx.lanl.gov/abs/1310.0378}{{\tt arXiv:1310.0378}}.

\bibitem{Chirilli:2013kca}
G.~A. Chirilli and Y.~V. Kovchegov, {\it {Solution of the NLO BFKL Equation and
  a Strategy for Solving the All-Order BFKL Equation}},  {\em JHEP} {\bf 1306}
  (2013) 055, [\href{http://xxx.lanl.gov/abs/1305.1924}{{\tt
  arXiv:1305.1924}}].

\bibitem{Albacete:2009fh}
J.~L. Albacete, N.~Armesto, J.~G. Milhano, and C.~A. Salgado, {\it {Non-linear
  QCD meets data: A global analysis of lepton- proton scattering with running
  coupling BK evolution}},  {\em Phys. Rev.} {\bf D80} (2009) 034031,
  [\href{http://xxx.lanl.gov/abs/0902.1112}{{\tt arXiv:0902.1112}}].

\bibitem{Albacete:2010sy}
J.~L. Albacete, N.~Armesto, J.~G. Milhano, P.~Quiroga-Arias, and C.~A. Salgado,
  {\it {AAMQS: A non-linear QCD analysis of new HERA data at small-x including
  heavy quarks}},  {\em Eur. Phys. J.} {\bf C71} (2011) 1705,
  [\href{http://xxx.lanl.gov/abs/1012.4408}{{\tt arXiv:1012.4408}}].

\bibitem{Kovner:1995ja}
A.~Kovner, L.~D. McLerran, and H.~Weigert, {\it Gluon production from
  non{A}belian {W}eizsacker-{W}illiams fields in nucleus-nucleus collisions},
  {\em Phys. Rev.} {\bf D52} (1995) 6231--6237,
  [\href{http://xxx.lanl.gov/abs/hep-ph/9502289}{{\tt hep-ph/9502289}}].

\bibitem{Kovchegov:1997ke}
Y.~V. Kovchegov and D.~H. Rischke, {\it Classical gluon radiation in
  ultrarelativistic nucleus nucleus collisions},  {\em Phys. Rev.} {\bf C56}
  (1997) 1084--1094, [\href{http://xxx.lanl.gov/abs/hep-ph/9704201}{{\tt
  hep-ph/9704201}}].

\bibitem{Kovchegov:1998bi}
Y.~V. Kovchegov and A.~H. Mueller, {\it Gluon production in current nucleus and
  nucleon nucleus collisions in a quasi-classical approximation},  {\em Nucl.
  Phys.} {\bf B529} (1998) 451--479,
  [\href{http://xxx.lanl.gov/abs/hep-ph/9802440}{{\tt hep-ph/9802440}}].

\bibitem{Kovchegov:2000hz}
Y.~V. Kovchegov, {\it Classical initial conditions for ultrarelativistic heavy
  ion collisions},  {\em Nucl. Phys.} {\bf A692} (2001) 557--582,
  [\href{http://xxx.lanl.gov/abs/hep-ph/0011252}{{\tt hep-ph/0011252}}].

\bibitem{Balitsky:2004rr}
I.~Balitsky, {\it {Scattering of shock waves in QCD}},  {\em Phys. Rev.} {\bf
  D70} (2004) 114030, [\href{http://xxx.lanl.gov/abs/hep-ph/0409314}{{\tt
  hep-ph/0409314}}].

\bibitem{Blaizot:2010kh}
J.~P. Blaizot, T.~Lappi, and Y.~Mehtar-Tani, {\it {On the gluon spectrum in the
  glasma}},  {\em Nucl. Phys.} {\bf A846} (2010) 63--82,
  [\href{http://xxx.lanl.gov/abs/1005.0955}{{\tt arXiv:1005.0955}}].

\bibitem{Krasnitz:1999wc}
A.~Krasnitz and R.~Venugopalan, {\it The initial energy density of gluons
  produced in very high energy nuclear collisions},  {\em Phys. Rev. Lett.}
  {\bf 84} (2000) 4309--4312,
  [\href{http://xxx.lanl.gov/abs/hep-ph/9909203}{{\tt hep-ph/9909203}}].

\bibitem{Krasnitz:2003nv}
A.~Krasnitz, Y.~Nara, and R.~Venugopalan, {\it Probing a color glass condensate
  in high energy heavy ion collisions},  {\em Braz. J. Phys.} {\bf 33} (2003)
  223--230.

\bibitem{Lappi:2003bi}
T.~Lappi, {\it Production of gluons in the classical field model for heavy ion
  collisions},  {\em Phys. Rev.} {\bf C67} (2003) 054903,
  [\href{http://xxx.lanl.gov/abs/hep-ph/0303076}{{\tt hep-ph/0303076}}].

\bibitem{Gelis:2008sz}
F.~Gelis, T.~Lappi, and R.~Venugopalan, {\it {High energy factorization in
  nucleus-nucleus collisions. 3. Long range rapidity correlations}},  {\em
  Phys.Rev.} {\bf D79} (2009) 094017,
  [\href{http://xxx.lanl.gov/abs/0810.4829}{{\tt arXiv:0810.4829}}].

\bibitem{ALbacete:2010ad}
J.~L. Albacete and A.~Dumitru, {\it {A model for gluon production in heavy-ion
  collisions at the LHC with rcBK unintegrated gluon densities}},
  \href{http://xxx.lanl.gov/abs/1011.5161}{{\tt arXiv:1011.5161}}.

\bibitem{Kharzeev:2001gp}
D.~Kharzeev and E.~Levin, {\it {Manifestations of high density QCD in the first
  RHIC data}},  {\em Phys. Lett.} {\bf B523} (2001) 79--87,
  [\href{http://xxx.lanl.gov/abs/nucl-th/0108006}{{\tt nucl-th/0108006}}].

\bibitem{Kharzeev:2000ph}
D.~Kharzeev and M.~Nardi, {\it {Hadron production in nuclear collisions at RHIC
  and high density QCD}},  {\em Phys. Lett.} {\bf B507} (2001) 121--128,
  [\href{http://xxx.lanl.gov/abs/nucl-th/0012025}{{\tt nucl-th/0012025}}].

\bibitem{Kharzeev:2007zt}
D.~Kharzeev, E.~Levin, and M.~Nardi, {\it {Hadron multiplicities at the LHC}},
  \href{http://xxx.lanl.gov/abs/0707.0811}{{\tt arXiv:0707.0811}}.

\bibitem{Kovchegov:2001sc}
Y.~V. Kovchegov and K.~Tuchin, {\it Inclusive gluon production in DIS at high
  parton density},  {\em Phys. Rev.} {\bf D65} (2002) 074026,
  [\href{http://xxx.lanl.gov/abs/hep-ph/0111362}{{\tt hep-ph/0111362}}].

\bibitem{Kharzeev:2003wz}
D.~Kharzeev, Y.~V. Kovchegov, and K.~Tuchin, {\it Cronin effect and high-p(t)
  suppression in p a collisions},  {\em Phys. Rev.} {\bf D68} (2003) 094013,
  [\href{http://xxx.lanl.gov/abs/hep-ph/0307037}{{\tt hep-ph/0307037}}].

\bibitem{Accardi:2012qut}
A.~Accardi, J.~Albacete, M.~Anselmino, N.~Armesto, E.~Aschenauer, {\em
  et.~al.}, {\it {Electron Ion Collider: The Next QCD Frontier - Understanding
  the glue that binds us all}},  \href{http://xxx.lanl.gov/abs/1212.1701}{{\tt
  arXiv:1212.1701}}.

\end{thebibliography}

\providecommand{\href}[2]{#2}\begingroup\raggedright\endgroup

\end{document}